\documentstyle[12pt,epsfig]{article}

\input{epsf}
\newcommand{\bel}[1]{\begin{equation}\label{#1}}
\newcommand{\bal}[1]{\begin{eqnarray}\label{#1}}
\newcommand{\be}{\begin{equation}}
\newcommand{\ee}{\end{equation}}
\newcommand{\ba}{\begin{eqnarray}}
\newcommand{\ea}{\end{eqnarray}}

\newcommand{\bes}{\begin{equation*}}
\newcommand{\ees}{\end{equation*}}
\newcommand{\del}{\partial}

\newcommand {\wt}{\widetilde}

\begin{document}
\begin{titlepage}

\begin{center}
\vspace{0.4in} {\Large \bf
The behavior of the sextic coupling for the scalar field at
the intermediate and  strong coupling regime}\\
\vspace{.3in}
{\large\em Gino J.~A\~na\~nos
\\
 Instituto de F\'{\i}sica Te\'orica-IFT\\
  Universidade Estadual Paulista
   \\ Rua Pamplona
145, S\~ao Paulo, SP  01405-900  Brazil
 \\
 E-mail: gananos@ift.unesp.br }
\subsection*{Abstract}
\end{center}
We study the behavior of the renormalized sextic coupling at the intermediate and strong
coupling regime for  the $\varphi^4 $ theory defined in
$d=2$-dimension. We found a good agreement
with the results obtained by the field-theoretical renormalization-group in the
Ising limit. In this work we use the lattice regularization method.
%
%
\end{titlepage}
\newpage
\baselineskip .37in
\section {Introduction}
The scalar field  $(\varphi^4)_{d=2}$ theory is known to have an
interactive continuum limit. This theory is an excellent
theoretical laboratory in which nonperturbative methods can be
tested. For theories in less than four space-time dimensions can
offer interesting and complex behavior as well as tractability,
and for example the case of three space-time dimensions, they can
even be directly physical, describing various planar condensed
matter systems.
The $\varphi^4$ theory has been study in the regimen of the intermediate and strong
coupling analytically using the path integral representation of the generating
functional  of the Green's function. Two of these methods are the high temperature
series approach of Baker and Kincaid~\cite{Baker} and the strong coupling expansion of
Bender et al ~\cite{strong1,strong2}. Each of these methods is based in the replacing the
continuum by a Euclidean lattice and expanding the kinetic energy. For example
in the strong coupling expansion the method consists in expanding the Green's functions in the
inverse powers of the bare coupling. Of course the result of this expansion will
be better as the  bare coupling is strong enough. This is because the expansion requires
that the coupling constant has to be large to perform the analytic expansion of the path integral.
Another nonperturbative analytic method which was proved to be a powerful tool is the
the field theoretical renormalization group~\cite{Wilson}.

The first study of the $\varphi^4$ theory  in the regimen of
strong coupling using lattice Monte Carlo simulation was done by
Cooper, Freedman and Preston~\cite{strong3}. They showed that the
$\varphi^4$ theory has a non zero scattering amplitude in one and
two dimensions and the asymptotic value for the renormalized coupling
constant was in good agreement with the
high temperature series and strong coupling expansion method. Freedman et al~\cite{Freedman}
continued the study in three and four dimensions and they found
Monte Carlo evidence that the lattice $(\varphi^4)_{d=4}$ theory
has a noninteracting continuum limit consistent with the result of
renormalization group calculation~\cite{Wilson} and strong
coupling expansion~\cite{strong1,strong2}. In contrast the
$(\varphi^4)_{d=3}$ theory appears to have an interacting
continuum limit. In this paper we are interested in the higher
order renormalized coupling constant in particular the sextic
coupling constant $g_6$. We expect that in the limit of strong
coupling this quantity will approach the asymptotic limit. This
because the theory in the strong coupling limit can be considered
to be in the the critical region where fluctuations are so strong
that they complete screen out the initial interaction so that the
behavior of the system becomes universal. In two dimensions, Sokolov
and Orlov used renormalization group expansion and
Pad\'e-Borel-Leroy resumation technique to get
$g_6$~\cite{Sokolov:1998dq}. In the literature there are few works
available of the sextic coupling constant using the approach of
lattice Monte Carlo. The first work was performed by Wheter~\cite{Wheter} and later
by Tsypin~\cite{Tsypin}. So it seems
appropriate the nonperturbative study using the lattice Monte
Carlo technique~\cite{Montvay} to get not only the asymptotic
value of the higher order renormalized coupling constant but also
the behavior on the intermediate and strong regimen.
%
Here as usual we consider $ \hbar =c= 1$.
\section{The effective potential}
We consider here  the $\phi^4$ theory in d-dimensions
Euclidean space in the presence of a source $J$, which  the bare
action is given by:
$$
S[\phi,J]= \int d^dx \left[ {1\over 2}(\del_{\mu}\phi_0)^2+{1\over
2} m^2\phi_0^2+{g\over 4!}\phi_0^4-J_0\phi_0 \right].
\label{action}
$$
We introduce the vacuum persistent function $Z[J]$:
\be
Z[J]=\int D[\phi] e^{-S[\phi,J]}
\ee
and from this we obtain the connected n-point Green's functions $G_n(x_1 \dots x_n )$
\be
G_n(x_1 \dots x_n )=\left. \frac{\delta}{\delta J(x_1)}\dots \frac{\delta}{\delta J(x_n)}
\ln Z(J)\right|_{J=0}
\ee
The renormalized $\Gamma^{(N)}(0) $  proper $N$-point
Green's function is obtained as follows:
the wave-function renormalization is obtained from the two-point
Green's function from:
\be
Z^{-1}=\left.\frac{G_2(p^2)}{dp^2}\right|_{p^2=0}
\ee
and the renormalized mass $m^2_r$ is defined by
\be
m^2_r=\left.Z\,G_2^{-1}(p^2)\right|_{p^2=0}
\ee
The effective action  $\Gamma[\phi_c]$ is defined by a functional Legendre
transform of $W[J]=\ln Z[J]$:
\be
  \Gamma[\phi_c] =  \int d^dx\; \phi_c(x) J(x) - W[J] ,\
  \label {legendtr}
\ee

It is well known that the effective action is the generator
 of proper Green's functions and in particular we can write
 the effective action in the form
 \be \Gamma[\phi_c]=\sum_{n} \frac{1}{n}\int d^dx_1 \dots d^dx_n
 \Gamma^{(n)}(x_1,\dots,x_n)\phi_c(x_1) \dots \phi_c(x_n).
 \ee

 Here $\Gamma^{(n)}(x_1,\dots,x_n) $ is the proper $n$-point
Green's functions in presence of the source $J$.
We define the effective potential, $U(\phi)$, by
  \be
  \Gamma[\phi]= \int \; d^dx \; \left[ U(\phi)+\frac{1}{2}(\partial_{\mu} \phi)^2\;Z+...\right]
   \label {a5} \  .
  \ee
an the renormalized effective potential
\be
U_{r}=\sum_{n=1}^{\infty} \frac{\Gamma_r^{(2n)}(0)}{2n!} \;\phi_r^{2n}
\ee
where $\phi_r=\phi \, Z^{-1/2}$
and the renormalized $\Gamma^{(n)}(0) $  proper $n$-point
Green's functions are given by
\be
\Gamma^{(n)}_r(0) =Z^{n/2}\Gamma^{(n)}(0)  \; .
\ee
from here we see that $U_{r}$ is the generating function of one particle
irreducible renormalized Green's function at zero external momentum
on all legs.\\
The particular interest to us are the renormalized coupling constant $\wt \Gamma_r^{(4)}(0)$
and the renormalized sextic coupling constant $\wt \Gamma_r^{(6)}(0)$ which can be
expressed as follows:
\be
\wt \Gamma_r^{(4)}(0)=\left. -Z^2 (\wt G_2^{-1})^4 \wt G_4 \right|_{p=0}
\label{f4}
\ee
and
\be
\wt \Gamma_r^{(6)}(0)=\left. -Z^3 (\wt G_2^{-1})^6( \wt G_6-10
\wt G_4^2 \wt G_2^{-1}) \right|_{p=0}.
\label{f6}
\ee
The quantity that will be extracted from lattice Monte Carlo simulation
is the dimensionless renormalized constant $g_{2n}$ defined by,
\be
g_{2n}=\frac{\Gamma_r^{(2n)}(0)}{m_r^{2n-nd+d}(2n)!}.
\ee
For  $d=2$ this expression becomes
\be
g_{2n}=\frac{\Gamma_r^{(2n)}(0)}{m_r^2(2n)!}
\ee
and the effective potential divided by $m^2_r$ can be written,
\be
U_{r}(\phi_r)/m_r^2=\frac 12 \phi_r^2+g_4\phi_r^4+g_6\phi_r^6+\dots
\ee
\section{Simulations results}
For the Monte Carlo simulation of the lattice field theory
described by Eq.~(\ref{action}) we used the standard Metropolis
algorithm.  In order to avoid the trapping into meta stable states
due to the underlying Ising dynamics we use the cluster algorithm using
the embedded dynamics for $\phi^4$ theory, according to the action,
\cite{Brower:1989mt}.
\begin{equation}
\label{Ising} S_{\mathrm{Ising}}  =
 \mbox{ } -  \sum_x \sum_{\hat{\mu}}
\left| \phi(x+\hat{\mu}) \phi(x) \right| s(x+\hat{\mu}) s(x) \;,
\end{equation}
where $s(x) = \mathrm{sign}(\phi(x))$.
Statistical errors are evaluated taking into account the
autocorrelation time in the statistical sample generated by the
Monte Carlo simulation.
In the simulation we used the lattices $16^2$, $20^2$, $36^2$ and $64^2$.
To minimize finite size effects we choose $1<<1/(a\,m_r)<<L$.
For the purpose of numerical simulation it is appropriated to work with
equivalence Lagrange on the lattice
\be S[\phi]= 2\kappa \sum_{x,\mu} \phi(x)\phi(x+e_\mu)+
\sum_x \phi(x)^2+\sum_x\lambda
(\phi(x)^2-1)^2 \label {action3}.
 \ee
The parameters $m^2$ and $g^2$ are related to $\kappa$ and $\lambda$ as follows,
\be  m^2=\frac{1-2\lambda}{\kappa}-2d \;\;\;\;\;
g=\frac{6\lambda}{\kappa^2}\, .
\ee
We found the bar parameters  ($m^2$,$g$)  for the different lattices
by fixing the value of $\lambda$ and varying the parameter $\kappa$ until we find an
acceptable value of the renormalized mass within few percent. All the values of bare mass
were negative. Of course we can not have a phase
transition since we imposed the renormalized mass to the condition $1<<1/(a\,m_r)<<L$.
The renormalized mass $m_r$ was calculated using the expression
\be
a^{-2} \, m_r^{-2}=\left(\frac{L}{2\pi}\right)^2 \left[ \frac{<\tilde
\phi(0)^2>-<|\tilde \phi(p)|^2>}{<|\tilde \phi(p)|^2>} \right]
\ee
where here $\tilde \phi$ is the Fourier transform of $\phi$ and
$p=(\frac{2\pi}{L \, a},0)$ is the smallest available non-zero
momentum.

For purpose of calculating the values of  $g_4$ and $g_6$ from simulation
we use lattice version of eqs.(\ref{f4}) and (\ref{f6}),
\be
g_4(4!) =\frac{\wt \Gamma_r^{(4)}(0)}{m_r^2}=-m_r^2\frac{<\wt
\phi(0)^4>-3<\wt \phi(0)^2>^2}{<\wt \phi(0)^2>^2},
\label{g4}
\ee

\begin{eqnarray}
g_6 (6!)&=&\frac{\wt \Gamma_r^{(6)}(0)}{m_r^2}\nonumber \\
&=& 10(\frac{\Gamma_r^{(4)}(0)}{m_r^2})^2
-(m_r^2)^2\frac{<\wt \phi(0)^6>-15<\wt \phi(0)^4> <\wt
\phi(0)^2>+ 30<\wt \phi(0)^2>^3}{<\wt \phi(0)^2>^3},
\label{g6}
\end{eqnarray}
where  $< \wt \phi(0)>$
is the Fourier transform evaluated at zero momentum .
We notice that eqs.(\ref{g4}) and (\ref{g6})  are efficient in the intermediate
and strong coupling regime where the
statistical errors are reasonable contrarily
to the case of weak coupling.
In table 1. we present the results of simulations for various lattices size
 for  a limit of strong coupling.
 \begin{table}
 \centering
\begin{tabular}{|c|c|c|c|}
  \hline
  L &  $g_4$  & $g_6$ \\
  \hline
  $16$ & $0.60 \pm 0.013$ & $ 1.18 \pm 0.06 $\\
  $20$ & $0.60 \pm 0.011$ & $1.16 \pm 0.04$\\
  $36$& $0.59 \pm 0.012$ & $1.12 \pm 0.05 $\\
  $64$& $0.585 \pm 0.012$ & $1.112 \pm 0.04 $\\
  \hline
\end{tabular}
\caption{}\label{1}
\end{table}
The results in table 1. are consistent with the values
obtained by the field-theoretical renormalization-group in the Ising limit
$g_4=0.6125$ and $g_6=1.10$ \cite{Sokolov:1998dq}.
In Fig. 1. we present the results of simulation for
$g_6$  and $g_4$ for a large range of the bare coupling constant ($g \sim 15-3232$).
We observe both $g_{4}$ and $g_6$  approach  to asymptotic constant value as we
expected. From the figure also we can see that the variation of both quantities
are small considering the large variation on $g$.
\begin{figure}[htb]
\centerline{\epsfig{file=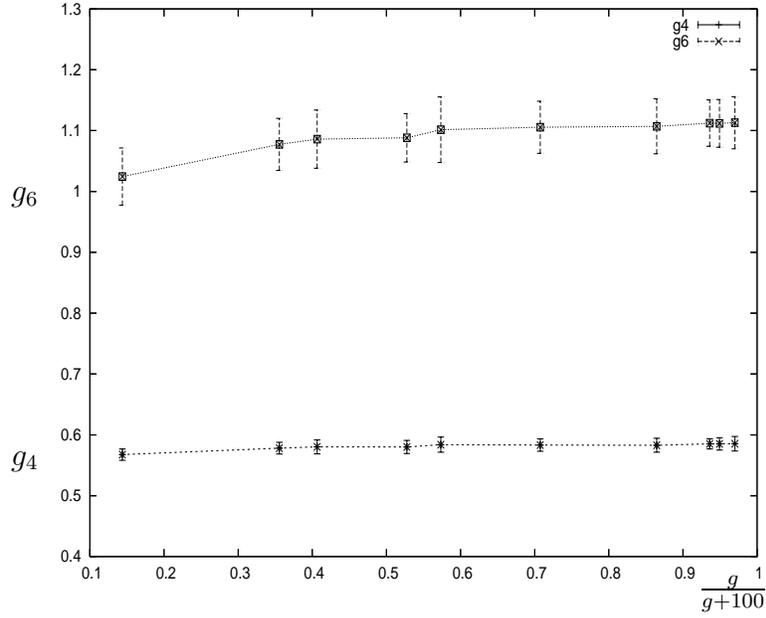,height=10cm,width=8cm,angle=-90}}
\caption{\small{The behavior of $g_4$ and $g_6$ with the bare coupling constant}}
\begin{picture}(120,80)(0,0)
\put(100,150){$g_4$}
\put(100,250){$g_6$}
\put(360,100){\small{$\frac{g}{g+100}$}}
\end{picture}
\end{figure}

\section{Conclusion}
In this paper we analyzed the behavior of the renormalized sextic coupling
for the theory $(\phi^4)_{d=2}$ on the lattice at strong coupling constant. We found
that this quantity tend to constant value that is in good agreement with
value obtained by using the field-theoretical renormalization-group in the Ising limit
at four loops. We also notice that the qualitative behavior of $g_6$ is similar to
$g_4$. We expect that this behavior is also similar for the case $d=3$, which will be
presented on future report.

\section{Acknowlegements}
This paper was supported by FAPESP under contract number 03/12271-7

\end{document}